\documentclass[a4paper, amsfonts, amssymb, amsmath, reprint, showkeys, nofootinbib, twoside, floatfix]{revtex4-1}
\usepackage[english]{babel}
\usepackage[utf8]{inputenc}
\usepackage[colorinlistoftodos, color=green!40, prependcaption]{todonotes}
\usepackage{amsthm}
\usepackage{mathtools}
\usepackage{physics}
\usepackage{xcolor}
\usepackage{graphicx}
\usepackage[left=23mm,right=13mm,top=35mm,columnsep=15pt]{geometry} 
\usepackage{adjustbox}
\usepackage{placeins}
\usepackage[T1]{fontenc}
\usepackage{lipsum}
\usepackage{csquotes}
\usepackage[pdftex, pdftitle={Article}, pdfauthor={Author}]{hyperref} 
\begin{document}
\title{Combined effects of Crab Dispersion and Momentum Dispersion in Colliders with Local Crab Crossing Scheme}

\author{Derong Xu}
\email{dxu@bnl.gov}
\author{Yun Luo}
    \affiliation{Brookhaven National Laboratory, Upton, New York 11973, USA}
\author{Yue Hao}
    \affiliation{Michigan State University, East Lansing, Michigan 48824, USA}

\date{\today} 

\begin{abstract}
  In this paper, we present the effects of linear transverse-longitudinal coupling on
  beam size at Interaction Point (IP) of a collider with local crab crossing scheme,  when time dependent transverse deflection (crab kicks) and dispersive orbit intertwine near IP. The analytic propagation formula and the closed orbit form of the crab dispersion and momentum dispersion are derived. The non-zero momentum dispersion
  at crab cavities and the non-ideal phase from crab cavities to IP are detailed with the derived propagation formula to predict the beam size distortion at IP with or
  without the beam-beam interaction.  The linear results are compared with nonlinear simulation using the weak-strong beam-beam code.
\end{abstract}



\maketitle

\section{Introduction} \label{sec:introduction}
A large crossing angle in the interaction region (IR) is necessary for fast separation of two colliding beams in ring-ring type colliders to achieve high collision rates, IR background minimization, and overall detector component and IR magnet arrangements.
Crab cavities, first proposed for linear colliders \cite{palmer1988energy}, 
can compensate for the geometrical luminosity loss induced by crossing angle. This idea was later expanded to include circular colliders \cite{oide1989beam}.
\par
The crab cavity generates a transverse kick, depending on
the longitudinal coordinate $z$ of a particle. Due to
symplecticity, the particle always receives an energy 
kick from the crab cavity as function of transverse offset $x$ 
simultaneously, as shown in Eq.~(\ref{eq:pxpz}).  
\begin{equation}
\begin{aligned}
    \Delta p_x&=-\lambda\sin\left(k_cz+\phi_c\right)/k_c\\
    \Delta \delta&=-\lambda\cos\left(k_cz+\phi_c\right)x
    \label{eq:pxpz}
\end{aligned}
\end{equation}
where $\Delta p_x$ and $\Delta \delta$ are horizontal and
energy kick from the crab cavity, $\lambda$ is the kick
strength normalized by the momentum of the reference particle,
$k_c$ and $\phi_c$ are the wave number and synchronous phase of the crab cavity.
\par
In crab crossing scheme, both colliding beams are tilted
by half crossing angle in $x-z$ plane to restore the
head-on collision. There are two configurations to accomplish
this: global or local schemes. 
In a global scheme, the crab cavity is placed at a 
particular location and the horizontal and longitudinal 
dynamics is coupled all over the ring.
In a local scheme, a pair of
crab cavities are installed at both sides of the IP. 
The upstream crab cavity tilts the beam in $x-z$ plane, 
and the downstream crab cavity rotates the beam back.
In the rest of the rings, both planes stay unaffected.
\par
\begin{figure}
    \centering
    \includegraphics[width=0.45\textwidth]{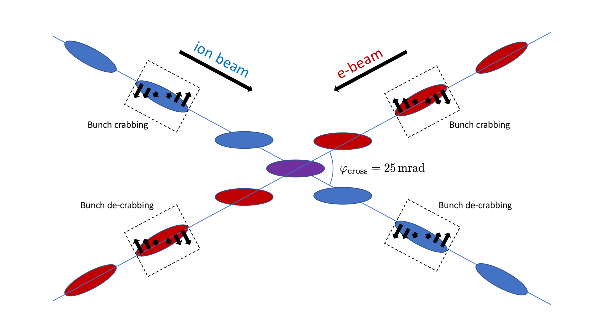}
    \caption{EIC local crabbing compensation scheme}
    \label{fig:EIC-local}
\end{figure}

The global scheme was first successfully implemented at the 
KEKB-factory \cite{abe2007compensationof}, where a world record 
luminosity of $2.1\times 10^{-34}~\mathrm{cm^{-2}s^{-1}}$ was 
obtained. The local scheme was also demonstrated for the hadron 
beam in CERN’s Super Proton Synchrotron (SPS) 
\cite{calaga2021first}. The Electron Ion
Collider (EIC) also adopts the local scheme to achieve the
desired luminosity ($1\times 10^{34}~\mathrm{cm}^{-2}\mathrm{s}^{-1}$) \cite{EIC-CDR}.
A schematic of the local crabbing compensation scheme
is shown in Fig.~\ref{fig:EIC-local} where two sets of crab cavities
are placed on both sides of IP for each ring.
\par
The single crab cavity dynamics in the global crabbing scheme
has been studied in detail. In the absence of longitudinal motion, the linear effect of crab cavities on the closed orbit is described by the concept of $z-$dependent 
dispersion \cite{sun2010crab}, which is referred as the crab dispersion 
throughout this study.
The linear transverse and longitudinal coupled motion due to crab
cavities is analyzed through the transfer matrix in 
\cite{huang2016coupled}. 
The synchro-betatron stop bands due to a single crab cavity are calculated in \cite{hoffstaetter2004synchrobetatron}.
The impact on the luminosity or the dynamical aperture is discussed in 
\cite{sun2009beam,funakoshi2014operational}.
\par
However, the crab cavity voltage in the global scheme depends on the linear beam optics which
is distorted by the beam-beam interaction. The crab dispersion all over the ring excites
various synchro-betatron resonances. From KEKB operation experiences, the global scheme
may be sensitive to the chromatic coupling and machine errors \cite{funakoshi2014operational}.
These can be avoided or mitigated in a local crabbing scheme 
as the crab dispersion is constrained within IR.
\par
In the ideal local crabbing scheme, the two crab cavities, located at the location with the betatron phase advance of $\pm\pi/2$ from IP, create desired crab dispersion "bump" between them.  The crab dispersion outside the crab cavity pair vanishes. Under this ideal assumption, the nonlinear $z-$ dependence from RF curvature and its impact of beam dynamics is described in \cite{xu2021synchrobetatron}. 

On the other hand, non-ideal crab-crossing setups also impact the dynamics of the colliding beams. The imperfections include the presence of  dispersion at crab cavities which are first discussed in \cite{chin1990effects}, and unmatched betatron phase advance between the crab cavity pair.  They break the closure of the crab dispersion bump and may cause degradation of beam quality and the luminosity. We present a theoretical treatment for the interplay of momentum and crab dispersion with these imperfections, then verify the predictions with the presence of the beam-beam effect in weak-strong simulations.

\par
This paper is organized as follows. Section \ref{sec:disperison} extends
the concept of crab dispersion and momentum dispersion to the 6-D phase space.
Section \ref{sec:nobeambeam} applies the theory to explain 
the effects of non-zero momentum dispersion at crab cavities, and non-ideal phase advance from crab cavities to IP.
Section \ref{sec:weak-strong} shows the results of combining the momentum/crab dispersion effects with beam-beam effect in a weak-strong simulation.
The conclusion is given in Sec. \ref{sec:conclusion}.
\section{Crab dispersion and momentum dispersion}\label{sec:disperison}
When the transverse coordinates $x$,$p_x$,$y$,
and $p_y$ are coupled with the 
longitudinal offset $z$ as well as the relative momentum deviation $\delta$,
neither $z$ nor $\delta$ is constant. In consequence,
the regular momentum dispersion is no longer well defined. 
We can instead define it as follows. 
\par
Let $\mathcal{M}$ be a canonical transformation
\begin{equation}
    \left(x,p_x,y,p_y,z,\delta\right)^\mathrm{T}
    =\mathcal{M}
    \left(\overline{x},\overline{p}_x,\overline{y},
    \overline{p}_y,\bar{z},\overline{\delta}\right)^\mathrm{T}
    \label{eq:transformation}
\end{equation}
where the superscript "T" denotes 
the transformation of a vector or a matrix.
In the new phase space of 
$\{\overline{x},\overline{p}_x,\overline{y},
    \overline{p}_y,\bar{z},\overline{\delta}\}$,
the longitudinal and transverse motion is decoupled.
Then the momentum dispersion and 
the crab dispersion are defined as
\begin{equation}
\boldsymbol{\eta}\equiv\frac{\partial \mathbf{X}}{\partial \overline{\delta}},
\qquad
\boldsymbol{\zeta}\equiv\frac{\partial \mathbf{X}}{\partial \overline{z}}
\label{eq:definitionOfCrabDispersion}
\end{equation}
where $\mathbf{X}$ is the abbreviation of 
$\left(x,p_x,y,p_y\right)^\mathrm{T}$. $\overline{z}$ and
$\overline{\delta}$ are connected by the longitudinal oscillation. As
a result, the two kinds of dispersion are also interchangeable.
\par
When the crab dispersion is not present, 
the transformation is well known \cite{chao2002lecture}
\begin{equation}
    \mathcal{M}_\eta=
    \left[\begin{matrix}
    \mathbf{1}_{4\times 4} & \mathbf{0}_{4\times 1} & \boldsymbol{\eta}\\
    -\left(J\boldsymbol{\eta}\right)^\mathrm{T} & 1 & 0\\
    \mathbf{0}_{1\times4} & 0 & 1
    \end{matrix}\right]
    \label{eq:M-eta}
\end{equation}
where $J$ is the 4-by-4 symplectic form matrix
\begin{equation}
    J=\left[\begin{matrix}
    0 & 1 & 0 & 0\\
    -1 & 0 & 0 & 0\\
    0 & 0 & 0 & 1\\
    0 & 0 & -1 & 0
    \end{matrix}\right]
    \label{eq:J4}
\end{equation}
Similarly, the transformation of the crab dispersion is
\begin{equation}
    \mathcal{M}_\zeta=
    \left[\begin{matrix}
    \mathbf{1}_{4\times 4} & \boldsymbol{\zeta} & \mathbf{0}_{4\times 1}\\
    \mathbf{0}_{1\times4} & 1 & 0\\
    \left(J\boldsymbol{\zeta}\right)^\mathrm{T} & 0 & 1\\
    \end{matrix}\right]
    \label{eq:M-zeta}
\end{equation}
When both kinds of dispersion are present, 
we can make a succession of
the two canonical transformations Eq.~(\ref{eq:M-eta}) 
and Eq.~(\ref{eq:M-zeta}),
\begin{equation}
\begin{aligned}
    \mathcal{M}&=\mathcal{M}_\zeta \mathcal{M}_\eta\\
    &=\left[\begin{matrix}
    \mathbf{1}_{4\times4}-\boldsymbol{\zeta}\left(J\boldsymbol{\eta}\right)^\mathrm{T} 
    & \boldsymbol{\zeta} & \boldsymbol{\eta} \\ 
    -\left(J\boldsymbol\eta\right)^\mathrm{T} &
    1 & 0\\
    \left(J\boldsymbol\zeta\right)^\mathrm{T} & 0 & 1+\left(J\boldsymbol{\zeta}\right)^\mathrm{T}\boldsymbol{\eta}
    \end{matrix}\right]
\end{aligned}
    \label{eq:M-zeta-eta}
\end{equation}
Substituting it back into 
Eq.~(\ref{eq:transformation}), it is straightforward 
to check that the transformation in Eq.~(\ref{eq:M-zeta-eta})
accommodates
the definition in Eq.~(\ref{eq:definitionOfCrabDispersion}),
\begin{equation}
    \mathbf{X}=M\overline{\mathbf{X}}+\boldsymbol\zeta \overline{z}+\boldsymbol\eta \overline\delta
\end{equation}
where $M$ is the 4-by-4 block of $\mathcal{M}$.
\par
The transformation $\mathcal{M}_\eta\mathcal{M}_\zeta$ 
also holds true for the definition in Eq.~(\ref{eq:definitionOfCrabDispersion}).
However, $\mathcal{M}_\zeta\mathcal{M}_\eta$ is a better choice
from the viewpoint of beam-beam study.
From Hirata \cite{hirata1995analysis}, the linear map for the 
Lorentz boost in the crab crossing scheme is
\begin{equation}
    \mathcal{L}\approx\left[
    \begin{matrix}
        1 & 0 & 0 & 0 & \theta_c & 0\\
        0 & 1 & 0 & 0 & 0 & 0\\
        0 & 0 & 1 & 0 & 0 & 0\\
        0 & 0 & 0 & 1 & 0 & 0\\
        0 & 0 & 0 & 0 & 1 & 0\\
        0 & -\theta_c & 0 & 0 & 0 & 1
    \end{matrix}
    \right]
    \label{eq:lorentzBoost}
\end{equation}
where $\theta_c$ is the half crossing angle, and the approximation 
$\theta_c\approx 0$ is used. The linear Lorentz boost 
$\mathcal{L}$ is literally a crab transformation with 
$\boldsymbol\zeta=\left(\theta_c,0,0,0\right)^\mathrm{T}$.
To provide an effective head-on collision, the crab dispersion
and the momentum dispersion are found to be
\begin{equation}
    \mathcal{L}\mathcal{M}=\mathbf{1}_{6\times6}
    \Longrightarrow
    \boldsymbol\zeta^*=(-\theta_c,0,0,0)^\mathrm{T},\ 
    \boldsymbol\eta^*=\mathbf{0}_{4\times1}
    \label{eq:headOnCollisionCondition}
\end{equation}
where the superscript symbol "*" denotes IP.
The property 
$\mathcal{M}_\zeta(\boldsymbol\zeta_1)\mathcal{M}_\zeta(\boldsymbol\zeta_2)=\mathcal{M}_\zeta(\boldsymbol\zeta_1+\boldsymbol\zeta_2)$
is used in Eq.~(\ref{eq:headOnCollisionCondition}).
\par
The linear motion through a section can be expressed via the
6-by-6 transfer matrix $\mathcal{R}$. In the phase space of
$\{\overline{x},\overline{p}_x,\overline{y},
    \overline{p}_y,\bar{z},\overline{\delta}\}$,
the transfer matrix will be
\begin{equation}
\overline{\mathcal{R}}=\mathcal{M}_2^{-1}\mathcal{R}\mathcal{M}_1
\qquad
\mathrm{or}
\qquad
\mathcal{M}_2\overline{\mathcal{R}}=\mathcal{R}\mathcal{M}_1
\label{eq:Rbar}
\end{equation}
where points $1$ and $2$ are the entrance and the exit of this section. According to the definition, the matrix $\overline{\mathcal{R}}$ is block diagonalized, i.e.
\begin{equation}
\begin{aligned}
    \overline{r}_{i5}&=0, & \overline{r}_{i6}&=0\\
    \overline{r}_{5i}&=0, & \overline{r}_{6i}&=0
\end{aligned}
    \label{eq:r1-4c5-6}
\end{equation}
where $i=1,2,3,4$, and $\overline{r}_{ij}$ are the matrix elements 
of $\overline{\mathcal{R}}$ at $i$th row, $j$th column.
\par
There are $8$ free variables in $\mathcal{M}_2$.
In the meantime, the number of independent constraints 
in Eq.~(\ref{eq:r1-4c5-6}) is also $8$. 
In principle, $\boldsymbol\zeta_2$ and 
$\boldsymbol\eta_2$ are determined by Eq.~(\ref{eq:Rbar}) and Eq.~(\ref{eq:r1-4c5-6}).
The propagation of 
the crab dispersion and momentum dispersion can be resolved.
\par
When the particle travels through a momentum dispersing section 
without any crab cavities or
RF cavities, the 6-by-6 transfer matrix will be
\begin{equation}
    \mathcal{R}_{\mathrm{dis}}=
    \left[
    \begin{matrix}
        R_d & \mathbf{0}_{4\times 1} & {\mathbf{D}}\\
        {\mathbf{B}}^\mathrm{T} & 1 & r_{56}\\
        \mathbf{0}_{1\times4} & 0 & 1
    \end{matrix}
    \right]
    \label{eq:RwithoutCavities}
\end{equation}
where $R_d$ is the 4-by-4 block, and $\mathbf{D}$ is the momentum dispersion generator. The symplectic condition requires
\begin{equation}
    R_d^\mathrm{T}JR_d=J,\qquad
    {\mathbf{B}}^\mathrm{T}=
    {\mathbf{D}}^\mathrm{T}JR_d
    \label{eq:RwithoutCavitiesSymplecticCondition}
\end{equation}
The block diagonalized matrix $\overline{\mathcal{R}}_\mathrm{dis}$ has a form of
\begin{equation}
    \overline{\mathcal{R}}_\mathrm{dis}=
    \left[
    \begin{matrix}
        \overline{R}_d & \mathbf{0}_{4\times 1} & \mathbf{0}_{4\times 1}\\
        \mathbf{0}_{1\times4} & \overline{r}_{55} & \overline{r}_{56}\\
        \mathbf{0}_{1\times4} & 0 & \overline{r}_{66}
    \end{matrix}
    \right]
    \label{eq:RbarWithoutCavities}
\end{equation}
with the symplectic constraint
\begin{equation}
    \overline{R}_d^\mathrm{T}J\overline{R}_d=J,
    \qquad
    \overline{r}_{55}\overline{r}_{66}=1
    \label{eq:RbarWithoutCavitiesSymplecticCondition}
\end{equation}
Substituting Eq.~(\ref{eq:M-zeta-eta}), Eq.~(\ref{eq:RwithoutCavities}) and Eq.~(\ref{eq:RbarWithoutCavities})
into Eq.~(\ref{eq:Rbar}), it follows that
\begin{equation}
\begin{gathered}
\overline{r}_{55}=
1+\mathbf{B}^{\mathrm{T}}\boldsymbol\zeta_1,\qquad
\boldsymbol\zeta_2=R_d\boldsymbol\zeta_1/\overline{r}_{55},\\
\overline{r}_{56}=\mathbf{B}^{\mathrm{T}}\boldsymbol\eta_1
+r_{56}\left[1+(J\boldsymbol\zeta_1)^\mathrm{T}\boldsymbol\eta_1\right],
\\
\boldsymbol\eta_2=\overline{r}_{55}\left\{R_d\boldsymbol\eta_1+\mathbf{D}\left[1+(J\boldsymbol\zeta_1)^\mathrm{T}\boldsymbol\eta_1\right]\right\}-
\overline{r}_{56}R_d\boldsymbol\zeta_1
\end{gathered}
\label{eq:propagationAlongSection}
\end{equation}
When $\boldsymbol\zeta_1=\mathbf{0}_{4\times1}$, the propagation turns into
\begin{equation}
    \boldsymbol\zeta_2=\mathbf{0}_{4\times1},
    \qquad
    \boldsymbol\eta_2=R_d\boldsymbol\eta_1+\mathbf{D}
\end{equation}
which is the same as the normal dispersion propagation.
\par
When the particle passes by a cavity-like element, the linear transfer matrix is
\begin{equation}
    \mathcal{R}_\mathrm{cav}=
    \left[
    \begin{matrix}
        R_c & {\mathbf{C}} & \mathbf{0}_{4\times 1} \\
        \mathbf{0}_{1\times4} & 1 & 0\\
        {\mathbf{A}}^\mathrm{T} & r_{65} & 1
    \end{matrix}
    \right]
    \label{eq:Rcavities}
\end{equation}
with the symplectic constraint
\begin{equation}
    R_c^\mathrm{T}JR_c=J,\qquad
    {\mathbf{A}}=
    R_c^\mathrm{T}J{\mathbf{C}}
    \label{eq:RcavitiesSymplecticCondition}
\end{equation}
where $R_c$ is also the 4-by-4 block, and $\mathbf{C}$ is the crab dispersion
generator.\par
The block diagonalized matrix $\overline{\mathcal{R}}_\mathrm{cav}$ is
\begin{equation}
    \overline{\mathcal{R}}_\mathrm{cav}=
    \left[
    \begin{matrix}
        \overline{R}_c & \mathbf{0}_{4\times 1} & \mathbf{0}_{4\times 1}\\
        \mathbf{0}_{1\times4} & 1 & 0\\
        \mathbf{0}_{1\times4} & \overline{r}_{65} & 1
    \end{matrix}
    \right]
    \label{eq:RbarCavities}
\end{equation}
with the symplectic constraint
\begin{equation}
    \overline{R}_c^\mathrm{T}J\overline{R}_c=J
    \label{eq:RbarCavitiesSymplecticCondition}
\end{equation}
Substituting Eq.~(\ref{eq:M-zeta-eta}), Eq.~(\ref{eq:Rcavities}) and Eq.~(\ref{eq:RbarCavities})
into Eq.~(\ref{eq:Rbar}), it follows that
\begin{equation}
\begin{gathered}
\boldsymbol\eta_2=R_c\boldsymbol\eta_1\\
\boldsymbol\zeta_2=R_c\boldsymbol\zeta_1+\mathbf{C}-
\frac{\left(r_{65}+\mathbf{A}^\mathrm{T}\boldsymbol\zeta_1\right)\boldsymbol\eta_2}%
{1+\left(J\boldsymbol\zeta_1\right)^\mathrm{T}\boldsymbol\eta_1+\mathbf{A}^\mathrm{T}\boldsymbol\eta_1}
\end{gathered}
\label{eq:propagationAlongCavity}
\end{equation}
When $\boldsymbol\eta_1=\mathbf{0}_{4\times1}$, the propagation turns into
\begin{equation}
    \boldsymbol\zeta_2=R_c\boldsymbol\zeta_1+\mathbf{C},
    \qquad
    \boldsymbol\eta_2=\mathbf{0}_{4\times1}
\end{equation}
\par
Eq.~(\ref{eq:RwithoutCavities}) and Eq.~(\ref{eq:Rcavities}) include most common
accelerator components in a real machine. For a one-turn map 
in which both
momentum dispersion generator $\mathbf{D}$ and crab dispersion generator 
$\mathbf{C}$ are present, the closed orbit condition is imposed 
on the two dispersion functions,
\begin{equation}
    \boldsymbol\eta_1=\boldsymbol\eta_2,
    \qquad
    \boldsymbol\zeta_1=\boldsymbol\zeta_2
    \label{eq:closedOrbitCondition}
\end{equation}
This fixed point problem can be resolved with the help of 
Edwards-Teng approach \cite{edwards20parameterization}.
\par
To use the Edwards-Teng approach, the discussion is limited within
the 4D phase space $\{x,p_x,z,\delta\}$. A general 4-by-4 transfer
matrix in terms of 2-by-2 blocks is
\begin{equation}
    R=\left[\begin{matrix}
        R_{xx} & R_{xz}\\
        R_{zx} & R_{zz}
    \end{matrix}\right]
    \label{eq:R44General}
\end{equation}
Following \cite{sagan1999linear}, $R$
is block diagonalized by 
\begin{equation}
    R=VUV^{-1}
    \label{eq:R44Diagonalized}
\end{equation}
with
\begin{equation}
    V=\left[\begin{matrix}
        g\mathbf{1}_{2\times2} & W\\
        -W^+ & g\mathbf{1}_{2\times2}
    \end{matrix}\right],
    \quad
    U=\left[\begin{matrix}
        U_{xx} & \mathbf{0}_{2\times2}\\
        \mathbf{0}_{2\times2} & U_{zz}
    \end{matrix}\right]
    \label{eq:R44GeneralDiagonalizedUV}
\end{equation}
where $W,U_{xx},U_{zz}$ are 2-by-2 blocks, $W^{+}$
the symplectic conjugate of $W$, and $g$ given by
\begin{equation}
    g^2+(\mathrm{det}~W)=1
    \label{eq:gWRelation}
\end{equation}
where "det" means taking the determinant. This paper does not include 
the concrete formula of $W$, which the reader can find in 
\cite{sagan1999linear}.
\par
With the closed orbit condition, 
the Eq.~(\ref{eq:Rbar}) turns into 
\begin{equation}
    \mathcal{R}=\mathcal{M}\overline{\mathcal{R}}\mathcal{M}^{-1}
\end{equation}
Comparing it with Eq.~(\ref{eq:R44Diagonalized}), $\mathcal{M}$ is related
to $V$ by a longitudinal scaling transformation, i.e.
\begin{equation}
    \mathcal{M}
    =\left[\begin{matrix}
        g\mathbf{1}_{2\times2} & W\\
        -W^+ & g\mathbf{1}_{2\times2}
    \end{matrix}\right]
    \cdot
    \left[\begin{matrix}
        P & \mathbf{0}_{2\times1} & \mathbf{0}_{2\times1}\\
        \mathbf{0}_{1\times2} & 1/g & 0\\
        \mathbf{0}_{1\times2} & 0 & g\\
    \end{matrix}\right]
    \label{eq:relationTeng}
\end{equation}
where $P$ is a 2-by-2 matrix.
From Eq.~(\ref{eq:relationTeng}),
the two kinds of closed orbit
dispersion are
\begin{equation}
    \left[\boldsymbol\zeta,\boldsymbol\eta\right]
    =W\left[\begin{matrix}
        1/g & 0\\
        0 & g
    \end{matrix}\right]
    \label{eq:closedOrbitDispersion}
\end{equation}
and the matrix $P$ follows
\begin{equation}
    W^+P=
    \left[\begin{matrix}
        g & 0\\
        0 & 1/g
    \end{matrix}\right]W^+
\end{equation}
\par
The crab dispersion and the momentum dispersion
at any point are related to the closed orbit form in Eq.~(\ref{eq:closedOrbitCondition})
by the propagation formulas Eq.~(\ref{eq:propagationAlongSection}) and 
Eq.~(\ref{eq:propagationAlongCavity}). 
Eq.~(\ref{eq:M-zeta-eta}) presents a technique to decouple the transverse and the longitudinal phase space following acknowledgment of the two types of dispersion. 
\section{Applications without beam-beam}\label{sec:nobeambeam}
\begin{figure}[!htbp]
    \centering
    \includegraphics[width=0.35\textwidth]{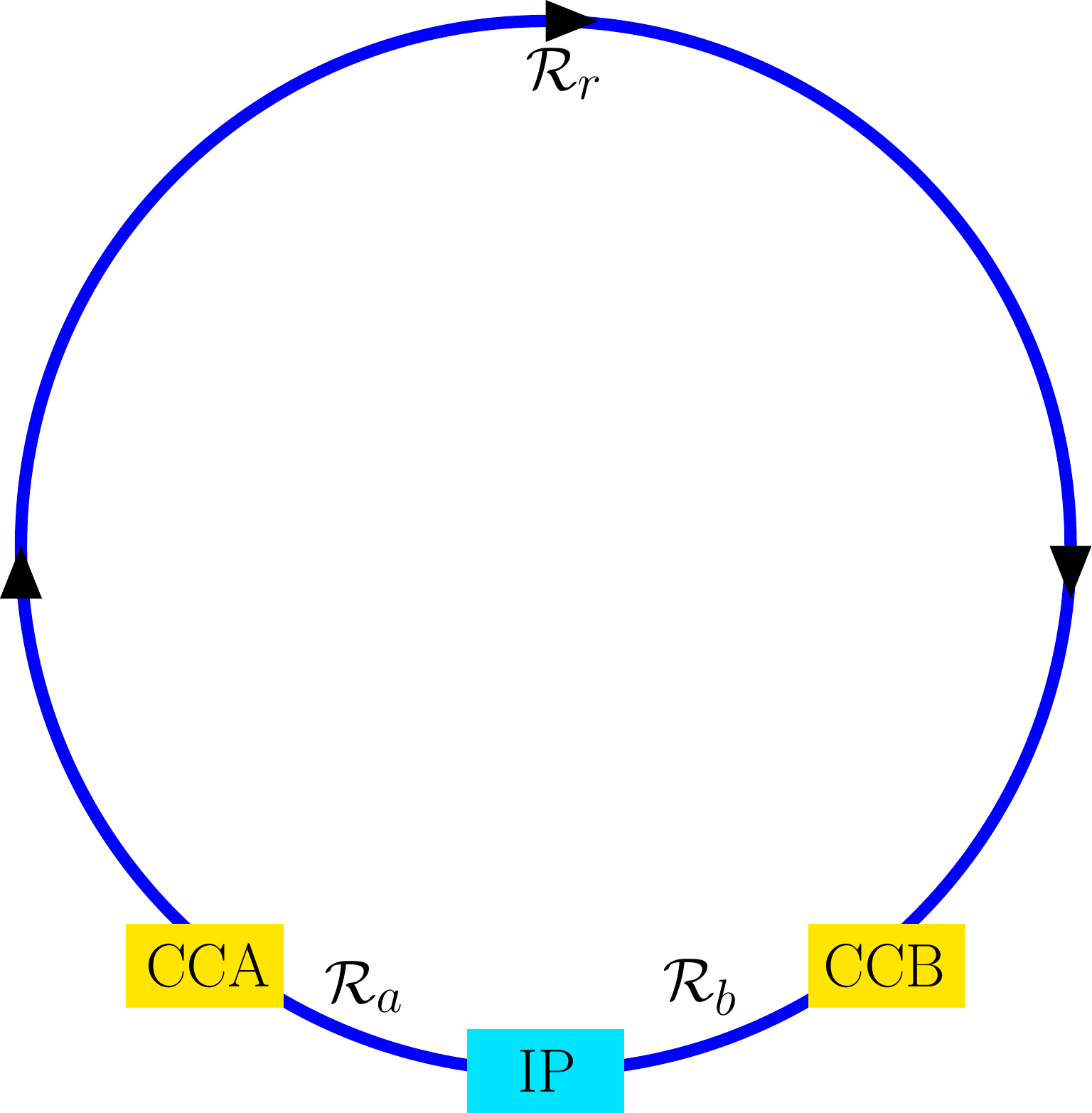}
    \caption{The local crabbing scheme in a storage ring. "CCB"
    stands for the crab cavity before IP, and
    $\mathcal{R}_b$ is the transfer matrix from CCB to IP.
    "CCA" stands for the crab cavity after IP, and
    $\mathcal{R}_a$ is the transfer matrix from IP to CCA.
    $\mathcal{R}_r$ is the transfer matrix from CCA to CCB.}
    \label{fig:cc6}
\end{figure}

In this section, the subscript "b" denotes before IP, 
whereas the subscript "a" denotes after IP. 
Without loss of generality, our discussion focuses in the
4D phase space $\{x,p_x,z,\delta\}$. The propagation of the crab dispersion
and the momentum dispersion don't involve the vertical plane. 
The lattice is assumed symmetrical around IP. 
The Crab Cavity Before IP (CCB) and the Crab Cavity After IP (CCA)
are placed at $\alpha_x=0$, as shown in Fig. \ref{fig:cc6}.
\subsection{Non-zero momentum dispersion at crab cavities}
\label{sec:momentumDispersionLeakage}
When the crab cavities are turned off, the momentum dispersion vanishes
at IP. Let the momentum dispersion at CCB be $\left(d,d'\right)^\mathrm{T}$. Then the transfer matrix
from CCB to IP is
\begin{equation}
    \mathcal{R}_b=\left[\begin{matrix}
        0 & \Lambda & 0 & -\Lambda d'\\
        -{1}/{\Lambda} & 0 & 0 & {d}/{\Lambda}\\
        d' & -d & 1 & r_{56}\\
        0 & 0 & 0 & 1
    \end{matrix}\right]
\end{equation}
where $\Lambda=\sqrt{\beta\beta^*}$, $\beta$ and $\beta^*$ are the horizontal beta functions at crab cavities and IP. 
From the symmetry of the lattice,
the momentum dispersion at CCA is $\left(d,-d'\right)^\mathrm{T}$, and
the transfer matrix from IP to CCA is
\begin{equation}
    \mathcal{R}_a=\left[\begin{matrix}
        0 & \Lambda & 0 & d\\
        -{1}/{\Lambda} & 0 & 0 & -d'\\
        -d/{\Lambda} & \Lambda d'  & 1 & r_{56}\\
        0 & 0 & 0 & 1
    \end{matrix}\right]
\end{equation}
\par
The periodic transfer matrix at IP should be
\begin{equation}
\begin{aligned}
    \mathcal{R}_{t1}&=\mathcal{R}_b\mathcal{R}_r\mathcal{R}_a\\
    &=\left[\begin{matrix}
        \cos\mu_x & \beta^*\sin\mu_x & 0 & 0\\
        -\frac{\sin\mu_x}{\beta^*} & \cos\mu_x & 0 & 0\\
        0 & 0 & \cos\mu_z & \frac{\sigma_z\sin\mu_z}{\sigma_\delta}\\
        0 & 0 & -\frac{\sigma_\delta\sin\mu_z}{\sigma_z} & \cos\mu_z
    \end{matrix}\right]
\end{aligned}
\end{equation}
where $\mathcal{R}_r$ the transfer matrix from CCA to CCB,
$\mu_x/\mu_z$ the periodic phase advance in horizontal/longitudinal
plane, $\sigma_z$ the RMS bunch length, and $\sigma_\delta$ 
the RMS momentum spread.
\par
When the crab cavities are turned on, from Eq.~(\ref{eq:pxpz}) 
the linear transfer matrix of CCB and CCA are
\begin{align}
    \mathcal{C}_b=\left[\begin{matrix}
        1 & 0 & 0 & 0\\
        0 & 1 & -\lambda_b & 0\\
        0 & 0 & 1 & 0\\
        -\lambda_b & 0 & 0 & 1
    \end{matrix}\right],\qquad
    \mathcal{C}_a=\left[\begin{matrix}
        1 & 0 & 0 & 0\\
        0 & 1 & -\lambda_a & 0\\
        0 & 0 & 1 & 0\\
        -\lambda_a & 0 & 0 & 1
    \end{matrix}\right]
\end{align}
where $\lambda_b$ and $\lambda_a$ are the strength of the crab cavity.
\par
Starting with
\begin{equation}
    \boldsymbol\zeta_0=(0,0)^{\mathrm{T}},\qquad \boldsymbol\eta_0=(0,0)^{\mathrm{T}}
\end{equation}
after transported to CCB by $\mathcal{R}_b^{-1}$,
defelected by $\mathcal{C}_b$, and transported back to IP by $\mathcal{R}_b$, the
crab dispersion and the momentum dispersion before collision read,
\begin{equation}
\begin{gathered}
    \boldsymbol\zeta_b=\left(-\frac{\Lambda\lambda_b}{1+\lambda_bd},\qquad 0\right)^\mathrm{T},\\
    \begin{aligned}
    \boldsymbol\eta_b=(1+\lambda_b d)\lambda_b d&\left(
        \Lambda d' ,\qquad
        -\frac{d}{\Lambda}
    \right)^\mathrm{T}
        \\
        +r_{56}&(1-\lambda_b d)
        \left(\Lambda\lambda_b,\qquad0\right)^\mathrm{T}
    \end{aligned}
\end{gathered}
\label{eq:dispersionBeforeCollision}
\end{equation}
Expanding Eq.~(\ref{eq:dispersionBeforeCollision}) to the first order of $\lambda_b$,
\begin{equation}
\begin{gathered}
\boldsymbol\zeta_b\approx \left(-\Lambda\lambda_b,0\right)^\mathrm{T},\\
\boldsymbol\eta_b\approx \Lambda\lambda_b\left(dd'+r_{56},-\frac{d^2}{\beta\beta^*}\right)^\mathrm{T}
\end{gathered}
\label{eq:dispersionBeforeCollision1stOrder}
\end{equation}
\par
\begin{figure}
    \centering
    \includegraphics[width=0.45\textwidth]{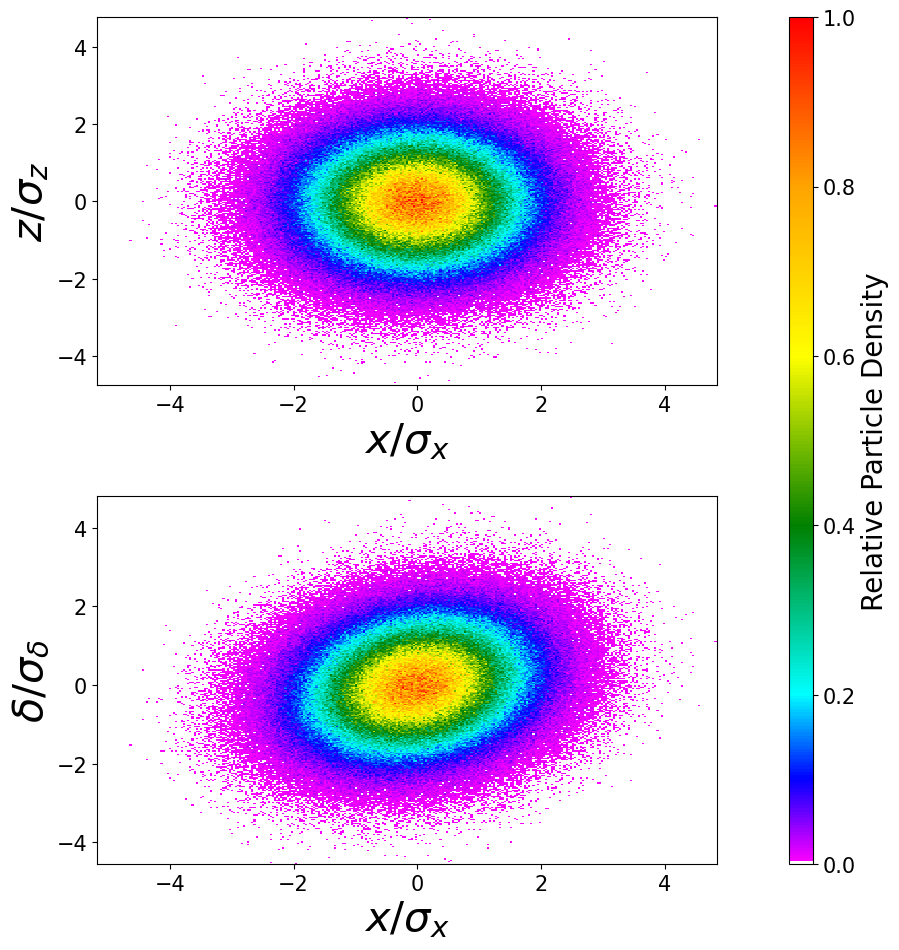}
    \caption{Beam distribution before collision in $x-z$ (top) and $x-\delta$ (bottom) plane. Both horizontal and vertical axes are normalized by RMS beam size. The
    dispersion at the crab cavity is $d=1~\mathrm{m}, d'=1$. The half crossing angle is $\theta_c=12.5~\mathrm{mrad}$. The $r_{56}$ element from CCB to IP is chosen as $r_{56}=2~\mathrm{m}$. The crab cavity strength is determined by $\lambda_b=\theta_c/\Lambda$. The horizontal
    beta functions at IP and the crab
    cavity are $\beta^*=0.5~\mathrm{m},\beta=200~\mathrm{m}$.}
    \label{fig:distribution}
\end{figure}

The Lorentz boost in Eq.~(\ref{eq:headOnCollisionCondition}) will cancel
the crab dispersion when $\lambda_b=\theta_c/\Lambda$. However,
the momentum dispersion doesn't vanish when $d\neq0$ or $d'\neq0$.
Therefore, the horizontal coordinate $x$ will depend on the momentum
spread $\delta$ in the head-on frame, as shown in Fig.~\ref{fig:distribution}.
\par
\begin{figure}
    \centering
    \includegraphics[width=0.45\textwidth]{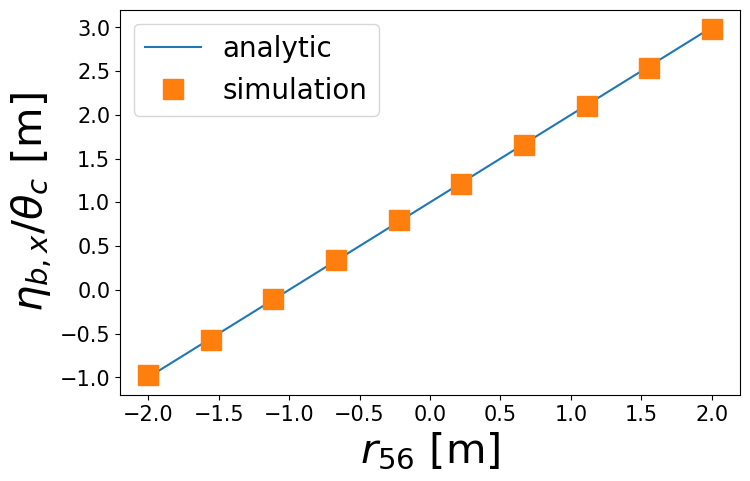}
    \caption{The horizontal momentum dispersion in the head-on frame versus the $r_{56}$ element from CCB to IP. The analytic
    line (blue) is obtained from Eq.~(\ref{eq:dispersionBeforeCollision1stOrder}), 
    and the simulation data (yellow) is from the statistics of the beam distribution.
    Other parameters are same as in Fig.~\ref{fig:distribution}.}
    \label{fig:comparison}
\end{figure}

Due to the non-zero dispersion, the transverse coordinates relate to
the momentum spread in the head-on frame by
\begin{equation}
    \mathbf{X}=\overline{\mathbf{X}}+\boldsymbol\eta_b\overline{\delta},
    \qquad
    \delta=\overline{\delta}
\end{equation}
The dispersion $\boldsymbol\eta_b$ can then be calculated from the second order 
moments as
\begin{equation}
\boldsymbol\eta_{b,x}=\frac{<x,\delta>}{\sigma_\delta^2},
\qquad
\boldsymbol\eta_{b,p_x}=\frac{<p_x,\delta>}{\sigma_\delta^2}
\label{eq:dispersionBeforeCollisionFromDistribution}
\end{equation}
where $<\cdot>$ denotes taking the average over the particle distribution.
Figure \ref{fig:comparison} compares the dispersion calculated from
the analytic formula Eq.~(\ref{eq:dispersionBeforeCollision1stOrder})
and from the beam distribution Eq.~(\ref{eq:dispersionBeforeCollisionFromDistribution}).
\par
Projecting the crab dispersion and the momentum dispersion at the other 
side back to IP, it follows
\begin{equation}
\begin{gathered}
    \boldsymbol\zeta_a=\left(\frac{\Lambda\lambda_a}{1+\lambda_a d},\qquad 0\right)^\mathrm{T},\\
    \begin{aligned}
    \boldsymbol\eta_a=(1+\lambda_a d)\lambda_a d&\left(
        \Lambda d' ,\qquad
        \frac{d}{\Lambda}
    \right)^\mathrm{T}
        \\
        +r_{56}&(1-\lambda_a d)
        \left(\Lambda\lambda_a,\qquad0\right)^\mathrm{T}
    \end{aligned}
\end{gathered}
\label{eq:dispersionAfterCollision}
\end{equation}
Expanding Eq.~(\ref{eq:dispersionAfterCollision}) to the first order of $\lambda_a$,
\begin{equation}
\begin{gathered}
\boldsymbol\zeta_a\approx \left(\Lambda\lambda_a,0\right)^\mathrm{T},\\
\boldsymbol\eta_a\approx\Lambda \lambda_a\left(dd'+r_{56},\frac{d^2}{\beta\beta^*}\right)^\mathrm{T}
\end{gathered}
\label{eq:dispersionAfterCollision1stOrder}
\end{equation}
Taking both sides into consideration, the crab dispersion 
can be closed when 
\begin{equation}
\lambda_b=\lambda_a\approx \theta_c/\Lambda
\end{equation}
Then the residual momentum dispersion is
\begin{equation}
    \boldsymbol\eta_a+\boldsymbol\eta_b\approx 2\theta_c\left(dd'+r_{56},0\right)^\mathrm{T}
\end{equation}
The leakage of the momentum dispersion will lead to the coupling between
the horizontal and longitudinal plane, and it is necessary to consider the closed
orbit form of the two types of distribution.
\par
Define
\begin{equation}
    k_\eta\equiv2\theta_c(dd'+r_{56})
    \label{eq:ketaDef}
\end{equation}
With both crab cavities on, the periodic transfer matrix at IP is
\begin{equation}
\begin{aligned}
    \mathcal{R}_{t2}&=(\mathcal{R}_b\mathcal{R}_r\mathcal{R}_a)(\mathcal{R}_a^{-1}\mathcal{C}_a\mathcal{R}_a)(\mathcal{R}_b\mathcal{C}_b\mathcal{R}_b^{-1})\\
    &\approx\mathcal{R}_{t1}\left[\begin{matrix}
    1 & 0 & 0 & k_\eta\\
    0 & 1 & 0 & 0\\
    0 & k_\eta & 1 & 0\\
    0 & 0 & 0 & 1
    \end{matrix}\right]\\
    &=\left[\begin{matrix}
        \cos\mu_x & \beta^*\sin\mu_x & 0 & k_\eta\cos\mu_x\\
        -\frac{\sin\mu_x}{\beta^*} & \cos\mu_x & 0 & -\frac{k_\eta\sin\mu_x}{\beta^*}\\
        0 & k_\eta\cos\mu_z & \cos\mu_z & \frac{\sigma_z\sin\mu_z}{\sigma_\delta}\\
        0 & -\frac{k_\eta\sigma_\delta\sin\mu_z}{\sigma_z}
        & -\frac{\sigma_\delta\sin\mu_z}{\sigma_z} & \cos\mu_z
    \end{matrix}\right]
\end{aligned}
\label{eq:Rt2}
\end{equation}
Following the procedure in \cite{sagan1999linear}, we define
\begin{equation}
\begin{gathered}
    H=\left[\begin{matrix}
        -\frac{k_\eta\sigma_\delta\sin\mu_z}{\sigma_z} & k_\eta(\cos\mu_x-\cos\mu_z)\\
        0 & -\frac{k_\eta\sin\mu_x}{\beta^*}
    \end{matrix}\right],\\
    g=\sqrt{\frac{1}{2}+\frac{1}{2}\sqrt{\frac{(\cos\mu_x-\cos\mu_z)^2}{(\cos\mu_x-\cos\mu_z)^2+\mathrm{det}~H}}}
\end{gathered}
\end{equation}
and then
\begin{equation}
    W=\frac{H}{2g\sqrt{(\cos\mu_x-\cos\mu_z)^2+\mathrm{det}~H}}
\end{equation}
\par
The stability criterion is
\begin{equation}
    (\cos\mu_x-\cos\mu_z)^2+\frac{k_\eta^2\sin\mu_x\sin\mu_z}{\beta^*\sigma_z/\sigma_\delta}>0
    \label{eq:stabilityCriterion}
\end{equation}
Similar to the betatron resonance, the sum resonance $\mu_x+\mu_z=0$ is dangerous,
while the motion on difference resonance $\mu_x-\mu_z=0$ is stable.
However, the longitudinal average action is usually much larger than 
the horizontal RMS emittance,
the coupling has to be weak enough to prevent the luminosity loss,
\begin{equation}
    g\approx 1,\qquad W\approx\frac{H}{2|\cos\mu_x-\cos\mu_z|}
\end{equation}
From Eq.~(\ref{eq:closedOrbitDispersion}), the two kinds of closed orbit
dispersion are
\begin{equation}
\begin{aligned}
    \zeta_{\mathrm{co},x}&=-\frac{k_\eta\sin\mu_z}{2|\cos\mu_x-\cos\mu_z|\sigma_z/\sigma_\delta}\\
    \eta_{\mathrm{co},x}&=\frac{1}{2}k_\eta\mathrm{sgn}(\cos\mu_x-\cos\mu_z)
\end{aligned}
\end{equation}
where 
\begin{equation}
    \mathrm{sgn}(x)=\left\{
    \begin{array}{lll}
    -1,& \mathrm{for} & x\leq0\\
    +1,& \mathrm{for} & x>0
    \end{array}
    \right.
\end{equation}
To prevent the horizontal beam size blows up, 
\begin{equation}
    |\zeta_{\mathrm{co},x}|\ll \frac{\sigma_x}{\sigma_z},
    \qquad
    |\eta_{\mathrm{co},x}|\ll \frac{\sigma_x}{\sigma_\delta}
\end{equation}
so that the constraints are given by
\begin{equation}
    \left|\frac{k_\eta}{2}\right|\ll \frac{\sigma_x|\cos\mu_x-\cos\mu_z|}{\sigma_\delta\sin\mu_z}
    \label{eq:constraintGivenByCrab}
\end{equation}
and
\begin{equation}
    \left|\frac{k_\eta}{2}\right|\ll \frac{\sigma_x}{\sigma_\delta}
    \label{eq:constraintGivenByMomentum}
\end{equation}
From Eq.~(\ref{eq:ketaDef}), $k_\eta$ and $\theta_c$ are within the same order of magnitude. As a result,
the constraint of Eq.~(\ref{eq:constraintGivenByMomentum})
is generally satisfied.
However, when $\mu_x$ is close to $\mu_z$, even if the stability criterion Eq.~(\ref{eq:stabilityCriterion}) holds,
the constraint of Eq.~(\ref{eq:constraintGivenByCrab}) may be broken.
In other words,
the leakage of momentum dispersion will result in a 
significant closed orbit crab dispersion, and the luminosity 
 will be reduced hereafter.
 \par
\begin{figure}
    \centering
    \includegraphics[width=0.45\textwidth]{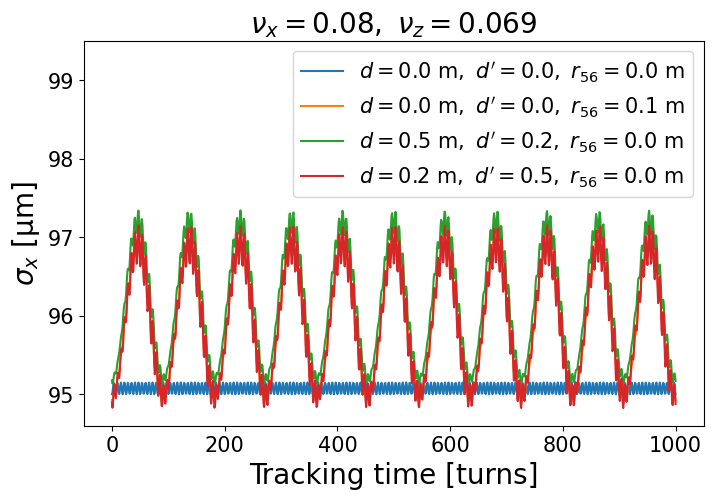}
    \includegraphics[width=0.45\textwidth]{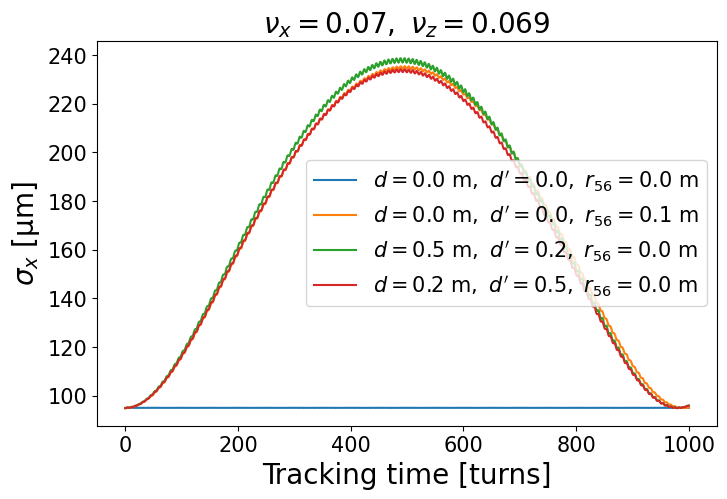}
    \caption{The horizontal beam size evolution due to the momentum distribution leakage. $\nu_x$ is the horizontal tune, $\nu_z$
    the longitudinal tune, and $(d,d')$ the horizontal momentum 
    dispersion at the crab cavities when the crab cavities are turned off. $r_{56}$ is the matrix element from CCB to IP, or from IP to CCA.}
    \label{fig:beamSizeEvolutionInMomentumLeakage}
\end{figure}

 The theory is verified by tracking. The macro particles are 
 randomly generated at IP from the Gaussian distribution,
 \begin{equation}
 \begin{aligned}
 \rho(x,p_x,&z,\delta)=\frac{1}{(2\pi)^2\sigma_{x}\sigma_{p_x}\sigma_{z}\sigma_{\delta}}\\
 &\times\mathrm{exp}\left(-\frac{x^2}{2\sigma_{x}^2}-\frac{p_x^2}{\sigma_{p_x}^2}
 -\frac{z^2}{2\sigma_{z}^2}-\frac{\delta^2}{2\sigma_{\delta}^2}\right)
 \end{aligned}
\label{eq:gaussianDistribution}
 \end{equation}
 The parameters are listed in Tab.~\ref{tab:initialSize}. The working point is chosen close to the difference resonance.
 \begin{table}
  \centering
  \caption{Initial beam size and crab cavity parameters in the tracking of dispersion leakage.}
   \begin{ruledtabular}
  \begin{tabular}{lcc}
    Parameter & Unit & Value \\
    \hline
    Horizontal size $\sigma_x$ & $\mathrm{\mu m}$ & $95.0$\\
    Horizontal divergence $\sigma_{p_x}$ & $\mathrm{\mu rad}$ & $211.0$\\
    Longitudinal size $\sigma_z$ & $\mathrm{cm}$ & $2.0$\\
    Momentum spread $\sigma_\delta$ & $10^{-4}$ & $5.5$\\
    Horizontal $\beta$ at IP & $\mathrm{m}$ & $0.45$\\
    Horizontal $\beta$ at crab cavity & $\mathrm{m}$ & $222.0$\\
    Crab cavity frequency & $\mathrm{MHz}$ & $200$ \\
    Crab cavity phase & $\mathrm{rad}$ & $0$ \\
    Half crossing angle $\theta_c$ & $\mathrm{mrad}$ & $12.5$\\
  \end{tabular}
  \end{ruledtabular}
  \label{tab:initialSize}
\end{table}
The sinusoidal kick Eq.~(\ref{eq:pxpz}) from the crab cavities is used during 
tracking. Figure~\ref{fig:beamSizeEvolutionInMomentumLeakage}
presents the beam size evolution caused by the momentum
distribution leakage. In our model, all elements are linear except
for the crab cavities. As a result, the beam envelope oscillates.
The oscillation amplitude is determined by $k_\eta$ in Eq.~(\ref{eq:ketaDef}) or $dd'+r_{56}$ to the first order,
which leads to the yellow, green and red curves overlap with each other.
When the horizontal tune $\nu_x$ is close to the longitudinal tune
$\nu_z$, the motion is still stable, but the 
envelope oscillation amplitude becomes much larger. 
If the coupling is weak enough, the oscillation frequency is determined by $|\nu_x-\nu_z|$ \cite{chao2002lecture}.
The envelope oscillation will
lead to the horizontal beam size blow-up when the non-linearity
is present, such as the beam-beam interaction, the chromaticity, or 
high-order magnetic fields.
\par
\begin{figure}
    \centering
    \includegraphics[width=0.45\textwidth]{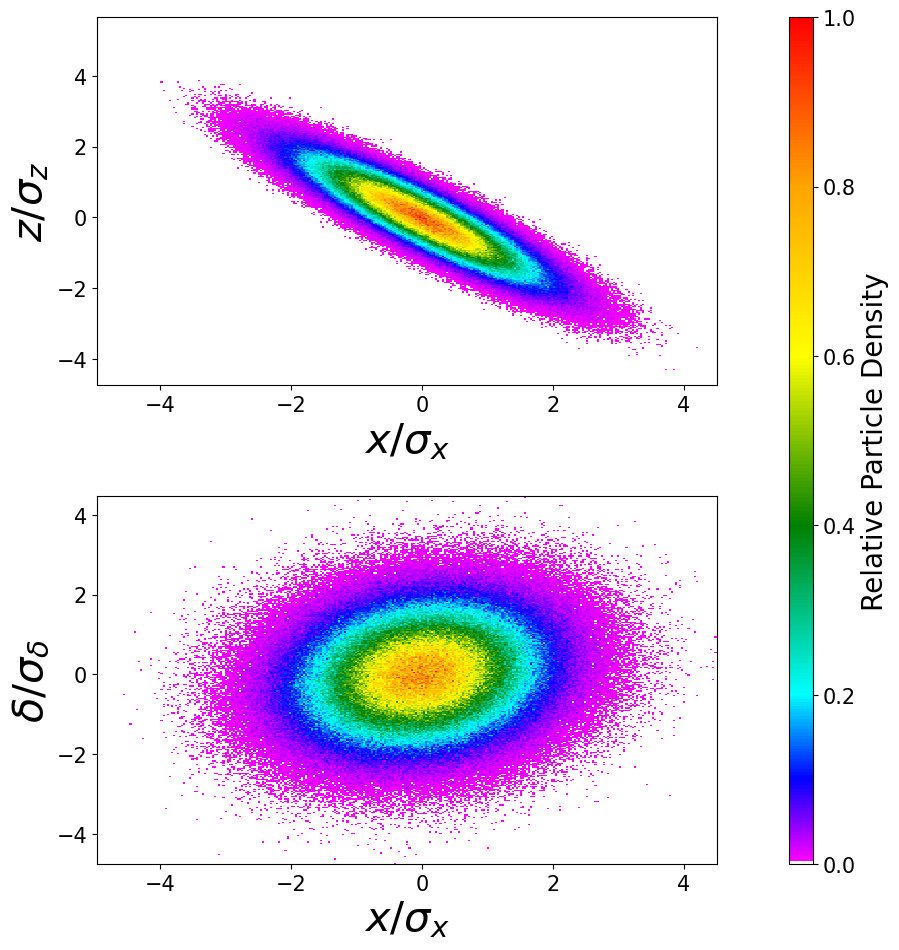}
    \caption{Beam distribution in $x-z$ (top) and $x-\delta$ (bottom) plane at $500$th turn for the
    green curve in the bottom of Fig.~\ref{fig:beamSizeEvolutionInMomentumLeakage}. Both horizontal and vertical axes are normalized by RMS beam size.}
    \label{fig:distributionMomentumLeakage}
\end{figure}

From Fig.~\ref{fig:beamSizeEvolutionInMomentumLeakage}, the
horizontal size reaches maximum at about $500$th turn.
Figure \ref{fig:distributionMomentumLeakage} shows the beam
distribution in $x-z$ and $x-\delta$ plane at that moment. It demonstrates
that the horizontal coordinate is substantially associated with the longitudinal
coordinate $z$, but weakly depending on the momentum spread $\delta$. It proves that
the closed orbit crab dispersion is significantly bigger than the momentum 
dispersion when the horizontal tune is close to the longitudinal tune.
\subsection{Non-ideal phase from crab cavities to IP}
\label{sec:crabDispersionLeakage}
The crab dispersion
from the crab cavities at both sides will cancel with each other when 
the horizontal phase advance from the crab cavity to IP is exactly $\pi/2$.
However, this is not always true in IR design. As a result, the crab 
dispersion will leak out of IR.
\par
Let $\Psi_b$ be the horizontal phase from CCB to IP, and $\Psi_a$ 
the horizontal phase from IP to CCA. 
The $\beta$ functions at both crab cavities are still assumed 
identical. We also omit the momentum dispersion in this section to simplify
our discussion. The transfer matrix between the crab cavities and IP
are given by,
\begin{equation}
\begin{aligned}
    \mathcal{R}_b&=\left[\begin{matrix}
        {\frac{\beta^*}{\Lambda}}\cos\Psi_b & \Lambda\sin\Psi_b
        & 0 & 0\\
        -\frac{\sin\Psi_b}{\Lambda} & {\frac{\Lambda}{\beta^*}}\cos\Psi_b & 0 & 0\\
        0 & 0 & 1 & 0\\
        0 & 0 & 0 & 1
    \end{matrix}\right],\\
    \mathcal{R}_a&=\left[\begin{matrix}
        {\frac{\Lambda}{\beta^*}}\cos\Psi_a & \Lambda\sin\Psi_a
        & 0 & 0\\
        -\frac{\sin\Psi_a}{\Lambda} & {\frac{\beta^*}{\Lambda}}\cos\Psi_a & 0 & 0\\
        0 & 0 & 1 & 0\\
        0 & 0 & 0 & 1
    \end{matrix}\right]
\end{aligned}
\end{equation}
\par
Following the same procedure in Sec.~\ref{sec:momentumDispersionLeakage},
the crab dispersion before collision is,
\begin{equation}
    \boldsymbol\zeta_b=-\Lambda\lambda_b\left(\sin\Psi_b,
    \frac{\cos\Psi_b}{\beta^*}\right)^\mathrm{T}
    \label{eq:zetabCrabDispersionLeakage}
\end{equation}
when $\Psi_b\neq\pi/2$, the second term in $\boldsymbol\zeta_{b}$ 
will not be equal to $0$. It will
introduce additional synchro-betatron resonance, and will degrade the beam-beam performance.\par
Projecting the crab dispersion from CCA back to IP,
\begin{equation}
    \boldsymbol\zeta_a=-\Lambda\lambda_a
    \left(-\sin\Psi_a,
    \frac{\cos\Psi_a}{\beta^*}\right)^\mathrm{T}
\end{equation}
It is easy to show that the residual crab dispersion vanishes
when
\begin{equation}
    \Psi_b+\Psi_a=\pi,
    \qquad
    \lambda_a=\lambda_b=\frac{\theta_c}{\Lambda\sin\Psi_b}
\end{equation}
There will be a leakage of crab dispersion when the total phase
$\Psi_b+\Psi_a$ deviates from $\pi$. 
\par
Let
\begin{equation}
    \lambda_b=\frac{\theta_c}{\Lambda\sin\Psi_b},
    \qquad
    \lambda_a=\frac{\theta_c}{\Lambda\sin\Psi_a},
    \label{eq:CCstrength}
\end{equation}
Then the leakage of the crab dispersion will be
\begin{equation}
    \boldsymbol\zeta_a+\boldsymbol\zeta_b=\theta_c\left(0,-\frac{\cot\Psi_a+\cot\Psi_b}{\beta^*}\right)^\mathrm{T}
\end{equation}
\par
Define
\begin{equation}
    k_\zeta\equiv -\theta_c\left(\frac{\cot\Psi_a+\cot\Psi_b}{\beta^*}\right)
    \label{eq:kzetaDef}
\end{equation}
With the crab dispersion leakage, the periodic transfer matrix at IP is
\begin{equation}
\begin{aligned}
    \mathcal{R}_{t3}&=(\mathcal{R}_b\mathcal{R}_r\mathcal{R}_a)(\mathcal{R}_a^{-1}\mathcal{C}_a\mathcal{R}_a)(\mathcal{R}_b\mathcal{C}_b\mathcal{R}_b^{-1})\\
    &=\mathcal{R}_{t1}\left[\begin{matrix}
    1 & 0 & 0 & 0\\
    0 & 1 & k_\zeta & 0\\
    0 & 0 & 1 & 0\\
    k_\zeta & 0 & 0 & 1
    \end{matrix}\right]\\
    &=\left[\begin{matrix}
        \cos\mu_x & \beta^*\sin\mu_x & \beta^*k_\zeta\sin\mu_x & 0\\
        -\frac{\sin\mu_x}{\beta^*} & \cos\mu_x & 
        {k_\zeta\cos\mu_x} & 0\\
        \frac{k_\zeta\sigma_z\sin\mu_z}{\sigma_\delta} & 0 & 
        \cos\mu_z & 
        \frac{\sigma_z\sin\mu_z}{\sigma_\delta}\\
        {k_\zeta\cos\mu_z} & 0
        & -\frac{\sigma_\delta\sin\mu_z}{\sigma_z} & \cos\mu_z
    \end{matrix}\right]
\end{aligned}
\end{equation}
\par
The stability criterion becomes
\begin{equation}
    (\cos\mu_x-\cos\mu_z)^2+k_\zeta^2\sin\mu_x\sin\mu_z\beta^*\left(\sigma_z/\sigma_\delta\right)>0
\end{equation}
Assuming the longitudinal-horizontal coupling is weak enough, the two kinds of closed orbit dispersion are
\begin{equation}
    \zeta_{\mathrm{co},x}=
    \frac{\beta^*k_\zeta\sin\mu_x}{2|\cos\mu_x-\cos\mu_z|},\qquad
    \eta_{\mathrm{co},x}=0
    \label{eq:closedOrbitDispersionFromCrabLeakage}
\end{equation}
Then a constraint is given by
\begin{equation}
    \left|\frac{k_\zeta}{2}\right|\ll 
    \frac{\sigma_x|\cos\mu_x-\cos\mu_z|}{\beta^*\sigma_z\sin\mu_x}
    \label{eq:zetaConstraintFromCrabLeakage}
\end{equation}
The bunch length $\sigma_z$ is usually much larger than
the transverse size $\sigma_x$. Accordingly, Eq.~(\ref{eq:zetaConstraintFromCrabLeakage}) places a
strict constraint on $k_\zeta$.
\par
\begin{figure}
    \centering
    \includegraphics[width=0.45\textwidth]{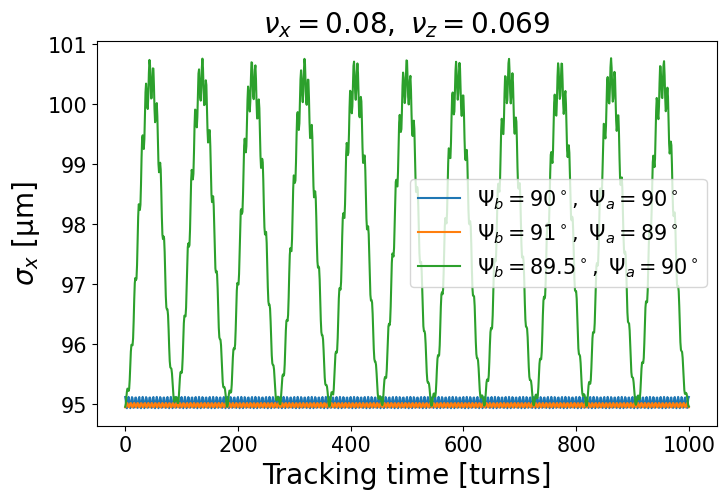}
    \includegraphics[width=0.45\textwidth]{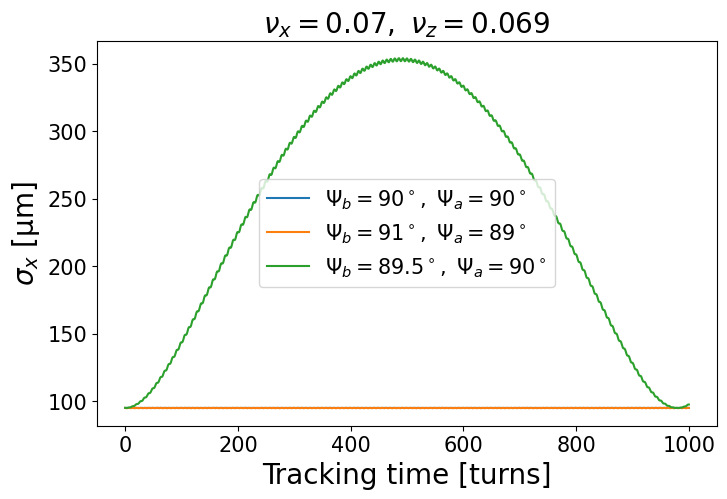}
    \caption{The horizontal beam size evolution due to the crab distribution leakage. $\nu_x$ is the horizontal tune, $\nu_z$
    the longitudinal tune. $\Psi_b$ is the horizontal phase advance from CCB to IP, and $\Psi_a$ is 
    the horizontal phase advance from IP to CCA.}
    \label{fig:beamSizeEvolutionInCrabLeakage}
\end{figure}
\begin{figure}
    \centering
    \includegraphics[width=0.45\textwidth]{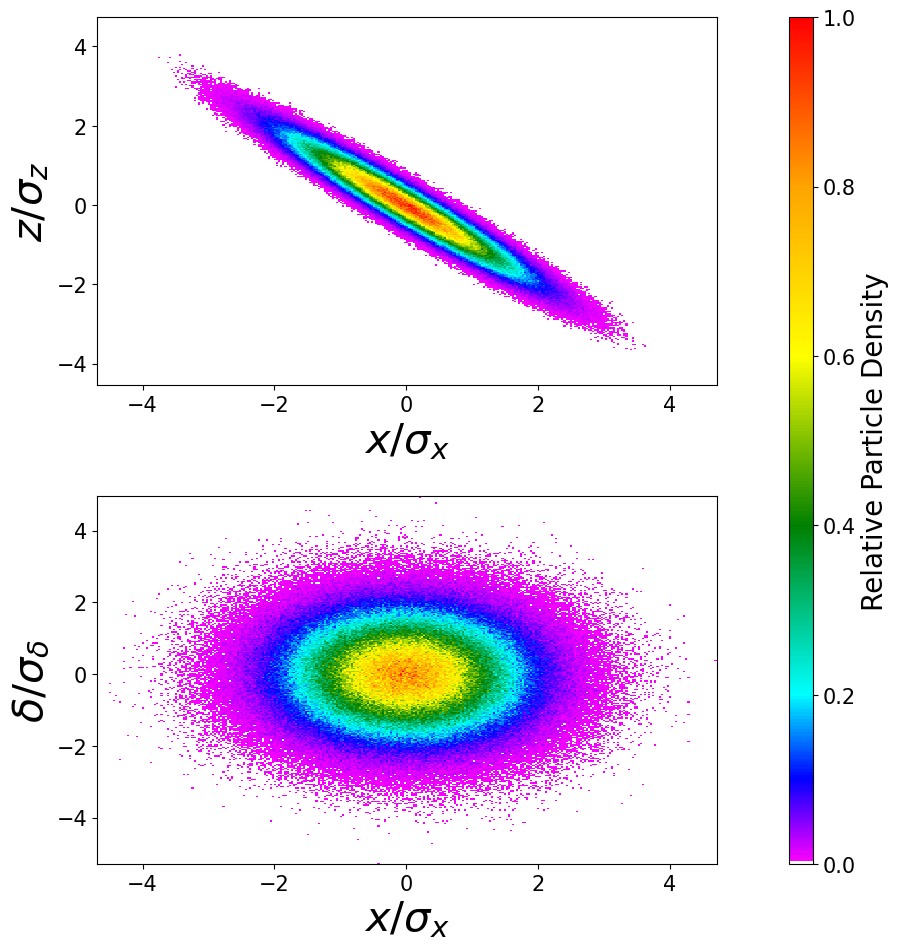}
    \caption{Beam distribution in $x-z$ (top) and $x-\delta$ (bottom) plane at $500$th turn for the
    green curve in the bottom of Fig.~\ref{fig:beamSizeEvolutionInCrabLeakage}. Both horizontal and vertical axes are normalized by RMS beam size.}
    \label{fig:distributionCrabLeakage}
\end{figure}

 Figure~\ref{fig:beamSizeEvolutionInCrabLeakage}
presents the beam size evolution caused by the crab
distribution leakage. The simulation parameters are listed in Tab.~\ref{tab:initialSize}. We can see that even $0.5^\circ$
deviation from $\pi$ driving a notable envelope oscillation
for the tunes $\nu_x=0.07, \nu_z=0.069$.
Figure \ref{fig:distributionCrabLeakage} shows the distribution in $x-z$ and $x-\delta$ plane when the horizontal envelope reaches maximum.
It turns out that it is the closed orbit crab
dispersion dominated the envelope oscillation, 
as predicted by Eq.~(\ref{eq:closedOrbitDispersionFromCrabLeakage}).
\section{Applications with beam-beam}\label{sec:weak-strong}
The leakage of the crab dispersion and momentum dispersion
will impose additional constraints on the lattice design.
Weak-strong simulation is a widely used approach in
beam-beam study \cite{papaphilippou1999weak,luo2012six}. 
In this part, we will investigate the influence of dispersion 
leakage on beam-beam performance using a self-written 
weak-strong code. 
\par
Table \ref{tab:EIC-CDR} presents the beam parameters used in 
the simulation to demonstrate the combined effects of crab dispersion
and momentum dispersion.
\begin{table}
  \centering
  \caption{Beam parameters in weak-strong simulation.
  The parameters come from EIC Conceptual Design Report \cite{EIC-CDR}.}
  \begin{ruledtabular}
  \begin{tabular}{lcc}
    Parameter & Proton & Electron \\
    \hline
    Circumference [$\mathrm{m}$] & \multicolumn{2}{c}{$3833.8$}  \\
    Energy [$\mathrm{GeV}$] & $275$ & $10$\\
    Particles per bunch [$10^{11}$] & $0.6881$ & $1.7203$\\
    Crossing angle [$\mathrm{mrad}$] & \multicolumn{2}{c}{$25.0$}\\
    Crab cavity frequency [$\mathrm{MHz}$] & $200.0$ & $400.0$ \\
    $\beta_x^*/\beta_y^*$ [$\mathrm{cm}$] & $80.0/7.20$ & $55.0/5.6$\\
    RMS emittance (H/V)[$\mathrm{nm}$] & $11.3/1.00$ & $20.0/1.30$\\
    RMS bunch size (H/V)[$\mathrm{\mu m}$] & {$95.0/8.50$} & $105/8.50$ \\
    RMS bunch length [$\mathrm{cm}$]& $6.0$ & $2.0$\\
    RMS energy spread [$10^{-4}$] & $6.6$ & $5.5$\\
    Transverse fractional tune (H/V) & $0.228/0.210$ & $0.08/0.06$\\
    Synchrotron tune  & $0.010$ & $0.069$\\
    Transverse damping time [turns] & $\infty$ & $4000$\\
    Longitudinal damping time [turns] & $\infty$ & $2000$\\
    Beam-beam parameter (H/V) & $0.009/0.009$ & $0.09/0.10$\\
  \end{tabular}
  \end{ruledtabular}
  \label{tab:EIC-CDR}
\end{table}
In the simulation, the ion beam is rigid with the horizontal centroid as \cite{xu2021synchrobetatron}
\begin{equation}
    x_{{i}}=-\theta_c\left[\frac{4}{3}\frac{\sin(k_{c,{i}}z)}{k_{c,{i}}}-\frac{1}{3}\frac{\sin(2k_{c,{i}}z)}{2k_{c,{i}}}-z\right]
\end{equation}
where $k_{c,i}$ is the wave number of the crab cavities in
the ion ring. A second order harmonic crab cavity is
used to flatten the ion bunch in the head-on frame.
The ion bunch is cut into multiple slices. Each slice is
represented by a 2D Gaussian distribution in $x-y$ plane.
\par
The weak electron beam are simulated by a number of macro particles. As in Sec. \ref{sec:momentumDispersionLeakage} and
Sec. \ref{sec:crabDispersionLeakage}, both the one-turn map 
and the betatron map from the crab cavities to IP are 
described by the linear transfer matrix. The crab cavity kick
follows Eq.~(\ref{eq:pxpz}). The beam-beam kick from a Gaussian
distribution is calculated with the Bassetti and Erskine formula \cite{bassetti1980closed}. The effects of radiation damping
and quantum excitation are represented by a lumped element \cite{qiang2002parallel}.
\par
\begin{figure}
    \centering
    \includegraphics[width=0.45\textwidth]{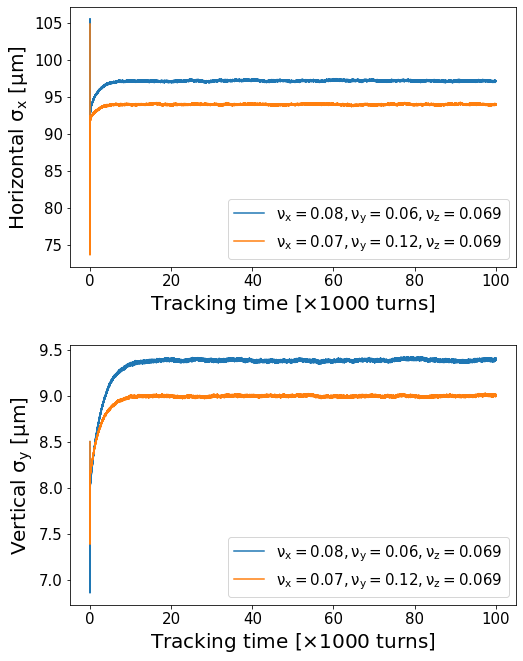}
    \caption{Weak-strong simulation results for the case without any dispersion leakage. The $\nu_x,\nu_y,\nu_z$ are horizontal, vertical and longitudinal tunes, respectively.}
    \label{fig:ws-ideal}
\end{figure}

Figure \ref{fig:ws-ideal} shows the beam size evolution 
without any dispersion leakage. Compared with the nominal
working point $\nu_x=0.08,\nu_y=0.06$ in the EIC CDR,
the new working point $\nu_x=0.07,\nu_y=0.12$ predicts smaller 
horizontal and vertical beam sizes after equilibrium. 
The horizontal size benefits from the smaller horizontal tune which
reduces the dynamical $\beta_x$ under beam-beam interaction \cite{furman1994beam}. The vertical size benefits from
a larger difference $|\nu_x-\nu_y|$ so that the new working
point moves away from the main diagonal line in the tune space.
From the viewpoint of beam-beam,
the new working point $\nu_x=0.07,\nu_y=0.12$ is a better choice.
\subsection{Non-zero momentum dispersion at crab cavities}
\begin{figure}
    \centering
    \includegraphics[width=0.45\textwidth]{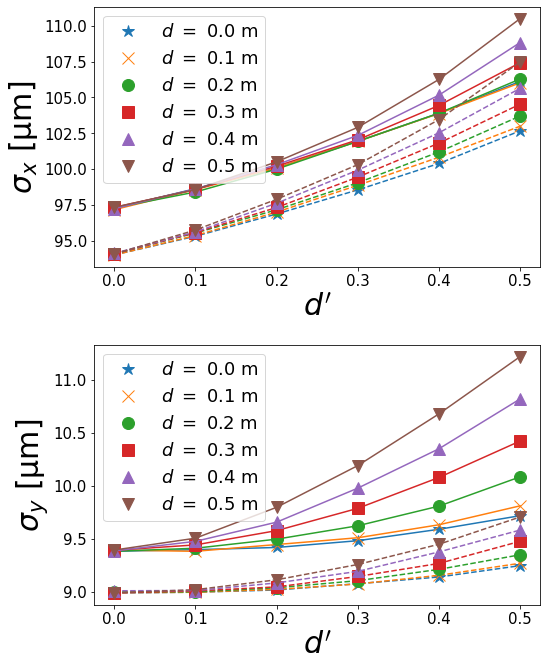}
    \caption{Weak-strong simulation results for different $d$ and $d'$. $(d,d')$ is the horizontal momentum dispersion at CCB when the crab cavities are turned off. The solid curves are for the
    working point $(0.08,0.06,0.069)$, while the dashed curves are
    for $(0.07,0.12,0.069)$. The horizontal size $\sigma_x$ and the 
    vertical size $\sigma_y$ are averaged
    from the last $1000$ turns.}
    \label{fig:ws-dispersion}
\end{figure}

Figure \ref{fig:ws-dispersion} presents the final beam sizes after 
equilibrium with
different momentum dispersion $d$ and $d'$ at both crab cavities.
The $r_{56}$ term from CCB (IP) to IP (CCA) is set to $0$ in all simulations.
Compared with the simulations without beam-beam interaction
in Sec. \ref{sec:momentumDispersionLeakage}, Figure \ref{fig:ws-dispersion} shows a quite different pattern.
The vertical size after equilibrium is also affected due to
the non-linearity from the beam-beam interaction. 
The horizontal blow-up is less severe even for the new working
point where the horizontal tune $\nu_x=0.07$ is quite close to the longitudinal tune $\nu_z=0.069$. The equilibrium size is 
mainly determined by $d'$ instead of $dd'+r_{56}$.
\par
The reason is that the horizontal tune and $\beta$ function are modified by
the beam-beam interaction. With the near axis approximation,
the beam-beam kick can be represented by a linear quadrupole
in the head-on frame,
\begin{equation}
    \mathcal{B}=\left[\begin{matrix}
        1 & 0 & 0 & 0\\
        -1/f_x & 1 & 0 & 0\\
        0 & 0 & 1 & 0\\
        0 & 0 & 0 & 1
    \end{matrix}\right]
    \label{eq:linearBB}
\end{equation}
where $f_x$ is the 
horizontal focal length, and can be expressed with the beam-beam parameter $\xi_x$ by
\begin{equation}
    \frac{1}{f_x}=\frac{4\pi\xi_x}{\beta^*}
\end{equation}
For simplicity, the vertical dynamics is not included here.
Back into the Frenet-Serret 
frame, the linear beam-beam transformation is given by,
\begin{equation}
    \mathcal{L}^{-1}\mathcal{B}\mathcal{L}=\left[\begin{matrix}
        1 & 0 & 0 & 0\\
        -1/f_x & 1 & -\theta_c/f_x & 0\\
        0 & 0 & 1 & 0\\
        -\theta_c/f_x & 0 & -\theta_c^2/f_x & 1
    \end{matrix}\right]
    \label{eq:linearBB2}
\end{equation}
where $\mathcal{L}$ is the linear Lorentz boost, as shown in Eq.~(\ref{eq:lorentzBoost}).
\par
Turning the crab cavities on, the total transfer matrix including
the crab system and the beam-beam interaction follows
\begin{equation}
\begin{aligned}
    &\mathcal{R}_{\mathrm{bb}}=(\mathcal{R}_a^{-1}\mathcal{C}_a\mathcal{R}_a)(\mathcal{L}^{-1}\mathcal{B}\mathcal{L})(\mathcal{R}_b\mathcal{C}_b\mathcal{R}_b^{-1})\\
    &\approx\left[\begin{matrix}
        1-a_x & a_x^2f_x & 0 & k_\eta\\
        -\frac{1}{f_x} & 1+a_x & 0 & -\frac{k_\eta}{2f_x}\\
        -\frac{k_\eta}{2f_x}\left(1+a_x\right) & k_\eta\left(1+a_x+\frac{a_x^2}{2}\right) & 1 & 0\\
        0 & 0 & 0 & 1
    \end{matrix}\right]
\end{aligned}
\end{equation}
where 
\begin{equation}
    a_x=\frac{2\Lambda\theta_cd'}{f_x}
\end{equation}
Then the periodic transfer matrix is
\begin{equation}
    \mathcal{R}_{t2,\mathrm{bb}}=\mathcal{R}_{t1}\mathcal{R}_{\mathrm{bb}}
\end{equation}
\par
\begin{figure}
    \centering
    \includegraphics[width=0.45\textwidth]{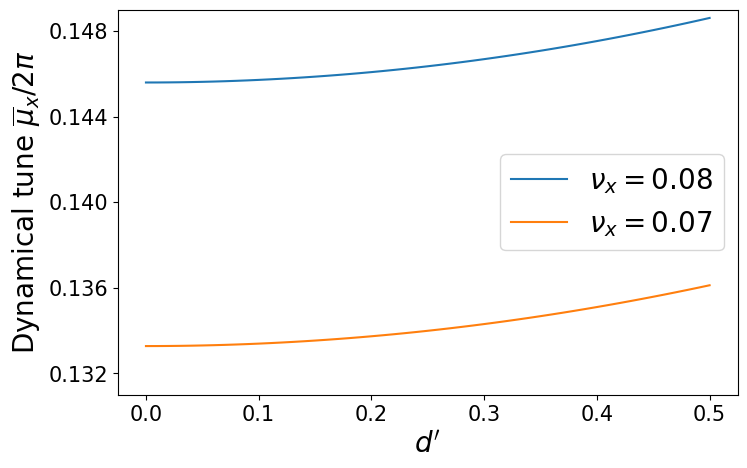}
    \caption{Dynamical tune in the presence of beam-beam interaction. $(d,d')$ is the horizontal momentum dispersion at CCB when the crab cavities are turned off.}
    \label{fig:dynamicalTuneDispersion}
\end{figure}
\begin{figure}
    \centering
    \includegraphics[width=0.45\textwidth]{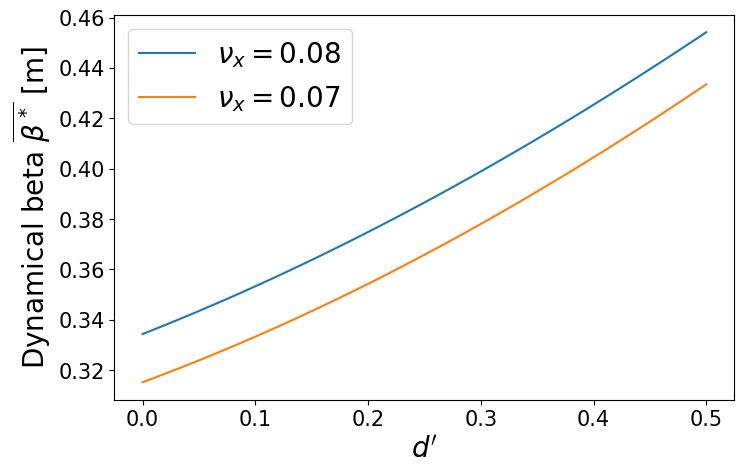}
    \caption{Dynamical beta function in the presence of beam-beam interaction. $(d,d')$ is the horizontal momentum dispersion at CCB when the crab cavities are turned off.}
    \label{fig:dynamicalBetaDispersion}
\end{figure}

Due to $k_\eta\propto \theta_c$, the longitudinal-horizontal coupling is still weak. 
The horizontal dynamic tune and $\beta$ are then given by
\begin{equation}
\begin{gathered}
    \cos\overline{\mu}_x = \cos\mu_x-\frac{1}{2}\left(
    \frac{\beta^*}{f_x}+\frac{a_x^2f_x}{\beta^*}
    \right)\sin\mu_x\\
    \overline{\beta^*}\sin\overline{\mu}_x=a_x^2f_x\cos\mu_x+\beta^*\left(1+a_x\right)\sin\mu_x
\end{gathered}
\label{eq:dynamicalTuneAndBeta}
\end{equation}
Figure \ref{fig:dynamicalTuneDispersion} shows the dynamical tune as a
function of $d'$. For both working points, the horizontal tune with
the beam-beam interaction is larger than $0.13$, which is far enough
away from the longitudinal tune $0.069$. Therefore, the closed orbit of
the momentum dispersion and crab dispersion are negligible.
Figure \ref{fig:dynamicalBetaDispersion} shows the dynamical beta as a
function of $d'$. The dynamical beta increases as $d'$ gets larger,
which explains why the horizontal size depends mainly on $d'$ instead
of $dd'+r_{56}$.
\par
It is worthwhile to mention that 
the dynamical beta is not the only source of beam size growth. The non-zero momentum dispersion $d$ or $d'$ at crab cavities will excite
higher-order synchro-betatron resonances through the nonlinear beam-beam
interaction.
\par
In summary, from the weak-strong simulation, when the dispersion satisfy
the constraints
\begin{equation}
    |d|<0.5~\mathrm{m},\qquad d'\sim 0
\end{equation}
the beam size growth caused by the momentum dispersion is small.
The closed orbit crab dispersion or momentum dispersion without beam-beam interaction are also negligible.
\subsection{Non-ideal phase from crab cavities to IP}
For symplecticity, we still omit the momentum dispersion here. 
Substituting Eq.~(\ref{eq:linearBB2}) into Eq.~(\ref{eq:propagationAlongCavity}), the crab dispersion
deflected by the beam-beam kick is
\begin{equation}
    \left[\begin{matrix}
        1 & 0\\
        -1/f_x & 1
    \end{matrix}\right]\boldsymbol\zeta_b
    +\left[\begin{matrix}
        0\\
        -\theta_c/f_x
    \end{matrix}\right]=\boldsymbol\zeta_b
\end{equation}
where $\boldsymbol\zeta_b$ takes the form of Eq.~(\ref{eq:zetabCrabDispersionLeakage}), and the crab cavity strength
is determined by Eq.~(\ref{eq:CCstrength}). Because the beam-beam kick has no effect on crab dispersion, the criteria Eq.~(\ref{eq:zetaConstraintFromCrabLeakage}) still holds true, with the exception that the horizontal phase must be replaced by the dynamical phase,
\begin{equation}
    \left|\frac{k_\zeta}{2}\right|\ll 
    \frac{\sigma_x|\cos\overline{\mu}_x-\cos\mu_z|}{\beta^*\sigma_z\sin{\mu}_x}
    \label{eq:zetaConstraintFromCrabLeakageBB}
\end{equation}
or specifically,
\begin{equation}
    |\cot\Psi_a+\cot\Psi_b|\leq 
    \frac{\sigma_x|\cos\overline{\mu}_x-\cos\mu_z|}{5\sigma_z\theta_c\sin{\mu}_x}
\end{equation}
where the dynamical phase $\overline{\mu}_x$ is determined by Eq.~(\ref{eq:dynamicalTuneAndBeta}), and the upper boundary is
set as $1/10$ in Eq.~(\ref{eq:zetaConstraintFromCrabLeakageBB}).
\par
\begin{figure}
    \centering
    \includegraphics[width=0.45\textwidth]{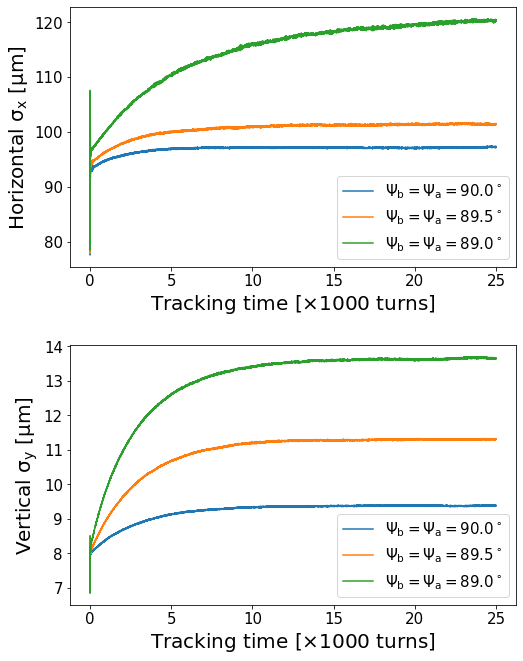}
    \caption{Weak-strong simulation results with non-ideal phase at
    the working point of $(0.08,0.06,0.069)$. $\Psi_b$ is the horizontal
    phase advance from CCB to IP, and $\Psi_a$ is the horizontal phase advance from IP to CCA.}
    \label{fig:ws-phase1}
\end{figure}
\begin{figure}
    \centering
    \includegraphics[width=0.45\textwidth]{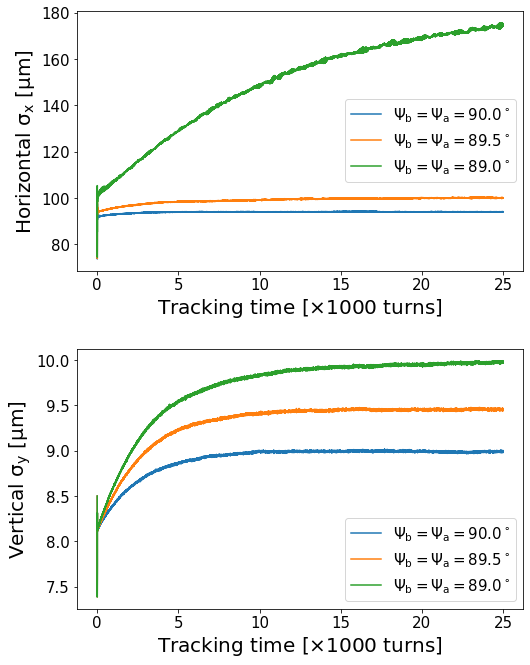}
    \caption{Weak-strong simulation results with non-ideal phase at
    the working point of $(0.07,0.12,0.069)$. $\Psi_b$ is the horizontal
    phase advance from CCB to IP, and $\Psi_a$ is the horizontal phase advance from IP to CCA.}
    \label{fig:ws-phase2}
\end{figure}

Let $\Psi_b=\Psi_a=\pi/2-\Delta\Psi$. Then 
the criterion becomes numerically,
\begin{equation}
\begin{aligned}
    |\Delta\Psi|&\leq 0.80^\circ\qquad\mathrm{when}\qquad\nu_x=0.08\\
    |\Delta\Psi|&\leq 0.58^\circ\qquad\mathrm{when}\qquad\nu_x=0.07
\end{aligned}
\label{eq:numericalCriterion}
\end{equation}
Figure \ref{fig:ws-phase1} and Fig. \ref{fig:ws-phase2} show
the weak-strong simulation results for different $\Delta\Psi$
at both working points. Both figures demonstrate that the constraint 
Eq.~(\ref{eq:numericalCriterion}) has to be satisfied. Otherwise,
the horizontal beam size will increase dramatically.
\par
However, the constraint of Eq.~(\ref{eq:numericalCriterion}) may be too strict to
meet in reality because of the compact layout of the IR. A possible
alternative is to move the crab cavities in one ring to the phase of
$3\pi/2$ or further.
\section{Conclusion} \label{sec:conclusion}
In this study, we extended the concept of crab dispersion and momentum dispersion in 
the presence of synchrotron motion. We derived the propagation law of the two types 
of dispersion traveling via common accelerator elements. Edward-Teng's block 
diagonalization technique was also used to find the closed orbit form of dispersions.
It enabled us to deduce the leakage of crab dispersion and momentum dispersion in 
the local crabbing scheme.
\par
This paper then investigated the momentum dispersion at the crab cavities and the non-ideal phase from the crab cavities to IP. The stability criterion was derived. A lattice requirement criterion was calculated using the weak horizontal-longitudinal coupling assumption. 
It turned out that the beam size at IP became sensitive to the leakage of dispersions when the horizontal tune was close to the longitudinal tune. The Monte Carlo simulations were
carried out to demonstrate the theoretical analysis.
\par
The beam-beam interaction was taken into consideration in the weak-strong simulations. 
It showed that the momentum dispersion at crab cavities had less impact on the beam size at IP because of the beam-beam tune shift. However, the phase advance from crab cavities to IP cannot stay too far from $\pi/2$. The numerical criteria of the electron ring lattice were given for the EIC beam parameters. The simulation results agreed with the criteria.

\section*{Acknowledgements} \label{sec:acknowledgements}
This work was supported by Department of Energy under Contract No. DE-AC02-98CH10886.
\bibliography{reference}
\end{document}